\documentclass[10pt,letterpaper,twocolumn]{article} %% two column, final layout
\usepackage{ol2}
\usepackage[draft,implicit=false]{hyperref}
\usepackage{amsmath}

\begin{document}
\twocolumn[
\title{Ultralong Faraday laser as an optical frequency standard}

\author{Chuanwen Zhu$^1$, Mo Chen$^1$, Xiaogang Zhang$^1$, Xiaobo Xue$^1$, Duo Pan$^1$, Jingbiao Chen$^{1,*}$ }

\address{
$^1$State Key Laboratory of Advanced Optical Communication Systems and Networks,\\
Institute of Quantum Electronics, School of Electronics Engineering and Computer Science,\\
Peking University, Beijing, 100871, China\\
$^*$Corresponding author: jbchen@pku.edu.cn
}

\begin{abstract}
In this letter, we introduce the concept and experimentally demonstrate an ultralong Faraday laser as an optical frequency standard in principle. The ultralong Faraday laser is based on the Faraday anomalous dispersion optical filter (FADOF) with ultra-narrow bandwidth and the ultralong fiber extended cavity of $800$ m. The ultra-narrow FADOF is based on atomic transition line of isotope $^{87}$Rb, which has an ultra-narrow bandwidth of $26.0$ MHz and a transmission of $23.6\%$ at $780$ nm. Fibers of length $150$ m and $800$ m are used as ultralong fiber extended cavities, which provide optical feedback and give extremely small FSR of $0.667$ MHz and $0.125$ MHz, respectively. The mechanism of the proposed ultralong Faraday laser is to combine FADOF's ultra-narrow bandwidth and ultralong cavity's small free spectral range to limit the lasing frequency within FADOF bandwidth covered by the semiconductor gain. The active lasing frequency of the ultralong Faraday laser is determined by the center frequency of FADOF transmission, which is corresponding to atomic transition $5^{2}S_{1/2},\ F=2 \ \rightarrow\ 5^{2}P_{3/2},\ F^{\prime}=2,3$ of isotope $^{87}$Rb. The Allan deviation of the fractional frequency of the ultralong Faraday laser output signal in 0.1 s -- 1 s measuring time is around  $3\times10^{-10}$.

\end{abstract}
]

Michael Faraday discovered rotation of the polarization plane of light in magnetic field, known as Faraday effect, in the year of 1845\cite{faraday}. 110 years after this discovery, with the enhancement of Faraday effect on atomic transition lines in atomic vapour observed by D. Macaluso and O. Corbino\cite{1898}, Y. \"{O}hman introduced and implemented an optical filter based on Faraday rotation of light in axial magnetic field\cite{ohman}, which is known as Faraday anomalous dispersion optical filter (FADOF). Not long after FADOF was introduced, it was added in laser cavity for laser mode selection and frequency stabilization\cite{fadoflock}. Based on that, FADOF was applied in semiconductor laser stabilization\cite{diode}\cite{optical}\cite{Afrequency} and linewidth reduction\cite{self}. Frequency stabilization with a combination of an electronic feedback from FM sideband spectroscopy and optical feedback has also been tested\cite{shevy1993linewidth}. In 2011, an extended cavity diode laser system with FADOF as frequency-selecting element and anti-reflection laser diode (ARLD) as gain medium was reported\cite{miao}, in which the output frequency is immune to the fluctuations of diode injection current or temperature.

FADOFs with different peak transmission wavelengths\cite{liu2011atomic}\cite{lingli} have potential applications in a variety of fields, like submarine communications, laser radar remote sensing systems\cite{Walther}, and hybrid continuous-variable/discrete-variable quantum optics\cite{zielinska2012ultranarrow}\cite{Zielinska}. The bandwidth of FADOFs can work on atomic transition lines with high transmission and narrow linewidth, and can even approach a bandwidth to the natural linewidth, such as 3.9 MHz bandwidth FADOF based on nonlinear spectroscopy of cesium\cite{wangyanfei}. The ultra-narrow bandwidth FADOFs can be applied to all-optical laser frequency stabilization, the reported accuracy approached 1.7 MHz\cite{xiaogang}.

In this letter, we are developing an ultralong Faraday laser as an optical frequency standard. The Faraday laser is a laser device that uses FADOF as frequency (longitudinal mode) selection element within laser cavity, its lasing frequency is predominantly determined by the FADOF transmission center. The ultralong Faraday laser works with the ultra-narrow FADOF and the ultralong cavity, the lasing frequency is determined by atomic transition line, this feature makes it an active optical frequency standard\cite{activeopticalclock}\cite{activefaraday} operating in good-cavity regime.  With only all-optical feedback of extended cavity, the laser is ``locked" in lasing configuration to atomic transition by using the ultra-narrow FADOF's transmission spectroscopy\cite{wangyanfei} as a frequency reference. Hence, the stable frequency output of the ultralong Faraday laser become an optical frequency standard, which is solely determined by the atomic transition.

We experimentally demonstrated the principle of the ultralong Faraday laser system, in which the frequency of ultra-narrow FADOF with Rb transition $5^{2}S_{1/2},\ F = 2 \ \rightarrow\ 5^{2}P_{3/2},\ F^{\prime} = 2,\ 3$ of isotope $^{87}$Rb at $780$ nm is used as quantum frequency reference. The ultralong fiber extended cavity is introduced, and the optical feedback is established by the coated high-reflection surface at the end of the fiber. The ultralong fiber cavity is used for mode selection via mode competition. Since the free spectral range (FSR) of the optical fiber cavity is given by $\frac{c}{2nL},$ an ultralong cavity length $L$ could provide ultra small FSR. If the FSR of the ultralong cavity is rather smaller compared to the bandwidth of the ultra-narrow FADOF, due to mode competition, the lasing frequency would always be near the center of FADOF's transmission peak, that is, the center frequency of atomic transition $5^{2}S_{1/2},\ F = 2 \ \rightarrow\ 5^{2}P_{3/2},\ F^{\prime} = 2,\ 3$ of $^{87}$Rb in our system. After all-optical feedback based on the ultralong cavity is estabilished, the ultralong Faraday laser provides a stable lasing frequency output centered at atomic transition with the help of ultra-narrow FADOF and ultralong fiber extended cavity.

In the first step, we measure the performance of the ultra-narrow FADOF\cite{wangyanfei} with $^{87}$Rb shown in the solid line box in Figure 1. The external cavity diode laser (ECDL) is tuned to the wavelength of crossover transition $5^{2}S_{1/2},\ F = 2 \ \rightarrow\ 5^{2}P_{3/2},\ F^{\prime} = 2,\ 3$ of $^{87}$Rb at $780$ nm. The PBS splits the ECDL's output laser beam into two beams. The one with larger power is used as the pumping beam (purple beam in Figure 1) and the one with smaller power is used as the probing beam (red beam in Figure 1). Part of the pumping beam split by the beam splitter is used for saturated absorption spectrum (SAS), which is used as the frequency reference to calibrate the bandwidth of the ultra-narrow FADOF transmission signal. The permanent magnets H$1$ and H$2$ create an axial magnetic field of $11$ G along the atomic vapor cell.

\begin{figure}[htbp]
 \centering
 \includegraphics[width=0.5\textwidth]{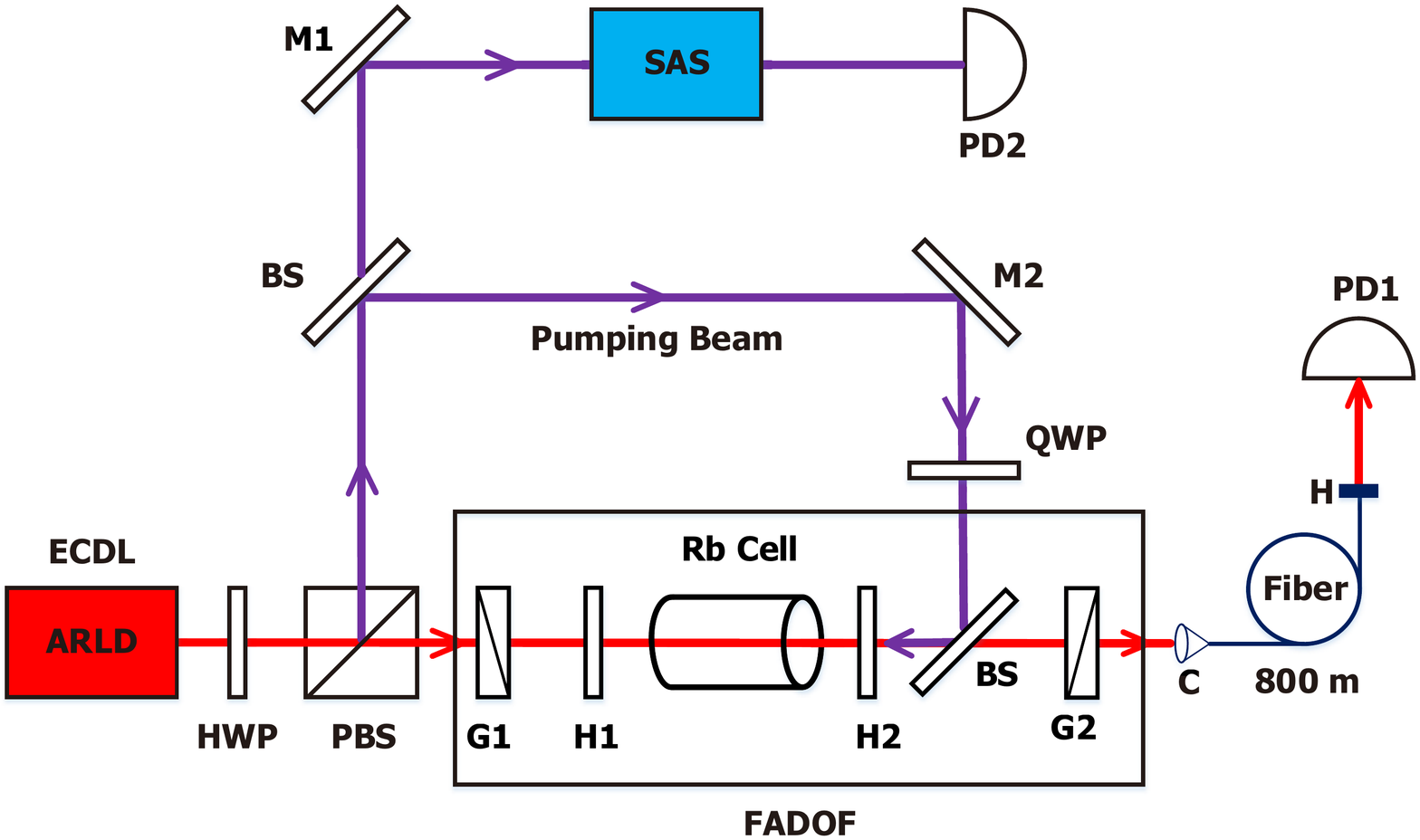}
 \caption{Experimental setup of ultralong Faraday laser. ECDL: external cavity diode laser, ARLD: anti-reflection laser diode,  SAS: saturated absorption spectrum, HWP: half-wave plate, QWP: quarter-wave plate, PBS: polarizing beam splitter, BS: beam splitter, M1/M2: 780 nm total reflection mirrors, G1/G2: A pair of orthogonally polarized Glan-Taylor prisms, H1/H2: permanent magnets, C: fiber coupler, H: high-reflection surface coated to fiber, PD1/PD2: photodetectors.}
 \label{schematics}
\end{figure}

The ultra-narrow FADOF is shown in the solid line box in Figure 1. By adjusting cell temperature and axial magnetic field, we manage to maximize an ultra-narrow bandwidth transmission peak of the FADOF at cell temperature of $68 ^{\circ}$C and magnetic field of $11$ G. From the photo detectors PD1 and PD2 in Figure 1, we can observe the FADOF transmission signal and SAS signal simultaneously. The result of the ultra-narrow FADOF signal is shown in Figure 2. We choose to operate the ultralong Faraday laser on the crossover transition $5^{2}S_{1/2},\ F = 2 \ \rightarrow\ 5^{2}P_{3/2},\ F^{\prime} = 2,\ 3$, which is indicated as co(2,3) in Figure 2. All the parameters of the ultra-narrow FADOF measured here are obtained without fiber extended cavity or optical feedback. The fitted bandwidth of this transmission peak is $26.0$ MHz calibrated by the SAS.

\begin{figure}[htbp]
 \centering
 \includegraphics[width=0.5\textwidth]{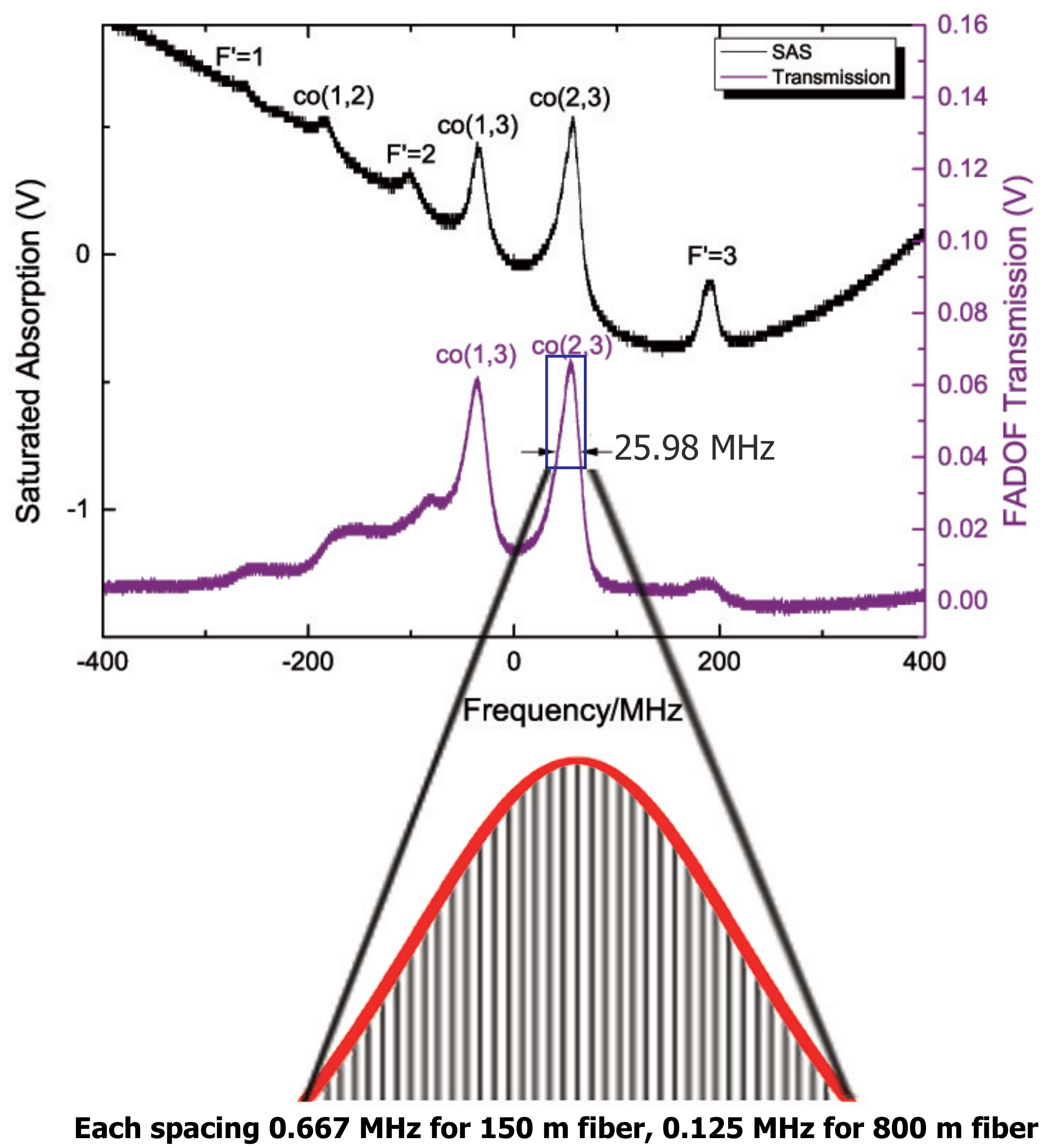}
 \caption{Ultra-narrow FADOF transmission signal (lower, purple) and SAS signal (upper, black) of Rb$^{87}$. All transitions are from ground state $5^{2}S_{1/2}$, F = 2 to excited state $5^{2}P_{3/2}, F^{\prime}=1, 2, 3$ and crossovers. The magnified part shows the transmission peak and FSR.}
 \label{FADOF}
\end{figure}

In the second step, we utilize the ultralong fiber cavity as shown in Figure 1, to set up the ultralong Faraday laser system. The ultralong fiber we fabricated for the system contains a coupler and a high-reflection surface at two ends, respectively. In this stage, the laser beam is coupled into the ultralong fiber by the coupler from one end, at the output end of the fiber, a high-reflection surface with $R=87\%$ is coated. The lengths of the ultralong fiber we used in the system are $150$ m and $800$ m, respectively. The high-reflection surface at the end of the fiber forms an ultralong fiber external cavity, which allows only resonating cavity modes exist. The light that carries the quantum transition information of the atoms in the ultra-narrow FADOF, whose frequency is selected and limited by the FADOF ultra-narrow bandwidth, is coupled into the fiber and reflected as feedback. The FSR of the fiber extended cavity is very small compared with the bandwidth of the ultra-narrow FADOF, as shown in Figure 2. Modes near the center of the ultra-narrow FADOF's bandwidth will compete (similar to general grating ECDL), resulting in only one lasing mode at the center.

As shown in Figure 2, the small FSR with long fiber will force the lasing frequency to stay at the center of the crossover transition of $5^{2}S_{1/2},\ F = 2 \ \rightarrow\ 5^{2}P_{3/2},\ F^{\prime} = 2,\ 3$ of $^{87}$Rb. Hence, we obtain a stable active lasing output on the atomic spectrum, which can also be treated as an optical frequency standard.

The stable laser output signal from PD1 in Figure 1 when using $150$ m long fiber in different
total measuring time is shown in Figure 3. To measure the performance of the ultralong Faraday laser as an optical frequency standard, the intensity fluctuation is transformed into the frequency fluctuation in a simplified way. By assuming the transmission profile being Lorentzian: $I\left(\Delta\right)=\frac{\left(\frac{\Gamma}{2}\right)^2}{\left(\frac{\Gamma}{2}\right)^2+\Delta^2}$, where $\Delta$ stands for frequency detuning and $\Gamma$ stands for linewidth of the profile. We derive the relationship between the fluctuation of transmitted intensity and the fluctuation of frequency near the center of transmission profile\cite{reid1982linewidth} to be $\delta\Delta=\frac{\Gamma}{2}\sqrt{|\delta I|}$. The Allan deviation of the transmitted intensity fluctuation is calculated, then transformed into the Allan deviation of frequency fluctuation by the formula above.

\begin{figure*}[htbp]
 \centering
 \includegraphics[width=\textwidth]{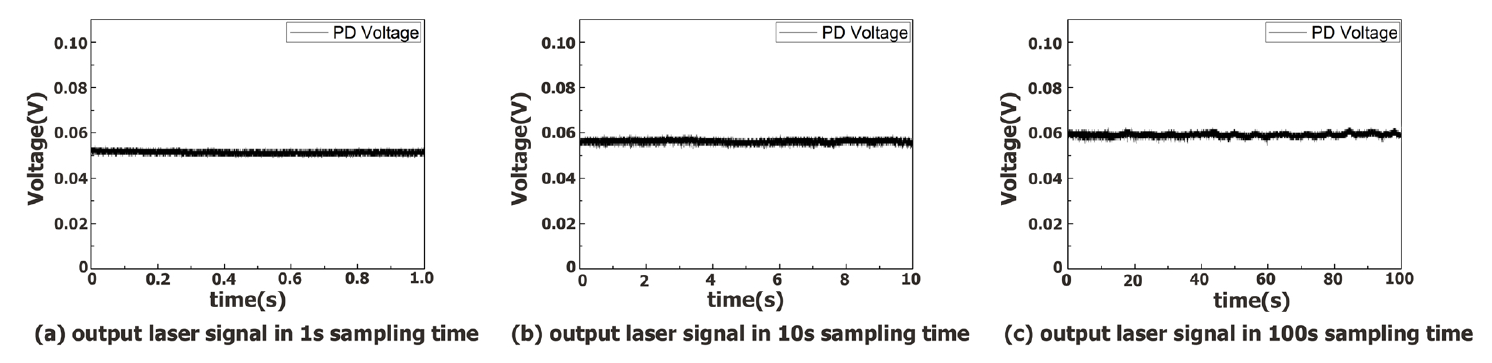}
 \label{lock}
 \caption{Output laser signal in 1s, 10s, 100s sampling time}
\end{figure*}
%Please note that this figure is not to be reduced to one-column width!!

The Allan deviation of the realized ultralong Faraday laser as an optical frequency standard when using fiber of $150$ m and $800$ m is shown in Figure 4. We can see that the fractional frequency stability can reach $10^{-9}$ order around $0.1$ s to $1$ s measuring time for both $150$ m and $800$ m long fiber. The $800$ m long fiber can give us a better short term stability since the laser Schawlow-Townes linewidth decreases with cavity length. For the long-term stability, the free drift of $800$ m fiber without length locking will contribute a limitation of $3.2\times10^{-10}$ with $1$ FSR of $125$ kHz. The Allan deviation for both lasers show a dropping in the last point. This is due to very limited locking time, which is only around $5$ s for $800$ m fiber, and 50--100 s for $150$ m fiber. In future work, antivibration and fiber (temperature and length) control technique will be applied to decrease the noise (and drift) from cavity. The output frequency has a drift range of FSR, which is determined by cavity length. With the technique of the fast scan of cavity length at a modulation frequency of $1$ kHz, i.e., averaging a number of cavity modes\cite{xiaogang}, the averaged frequency will be at the center of FADOF bandwidth.

\begin{figure}[htbp]
 \centering
 \includegraphics[width=0.5\textwidth]{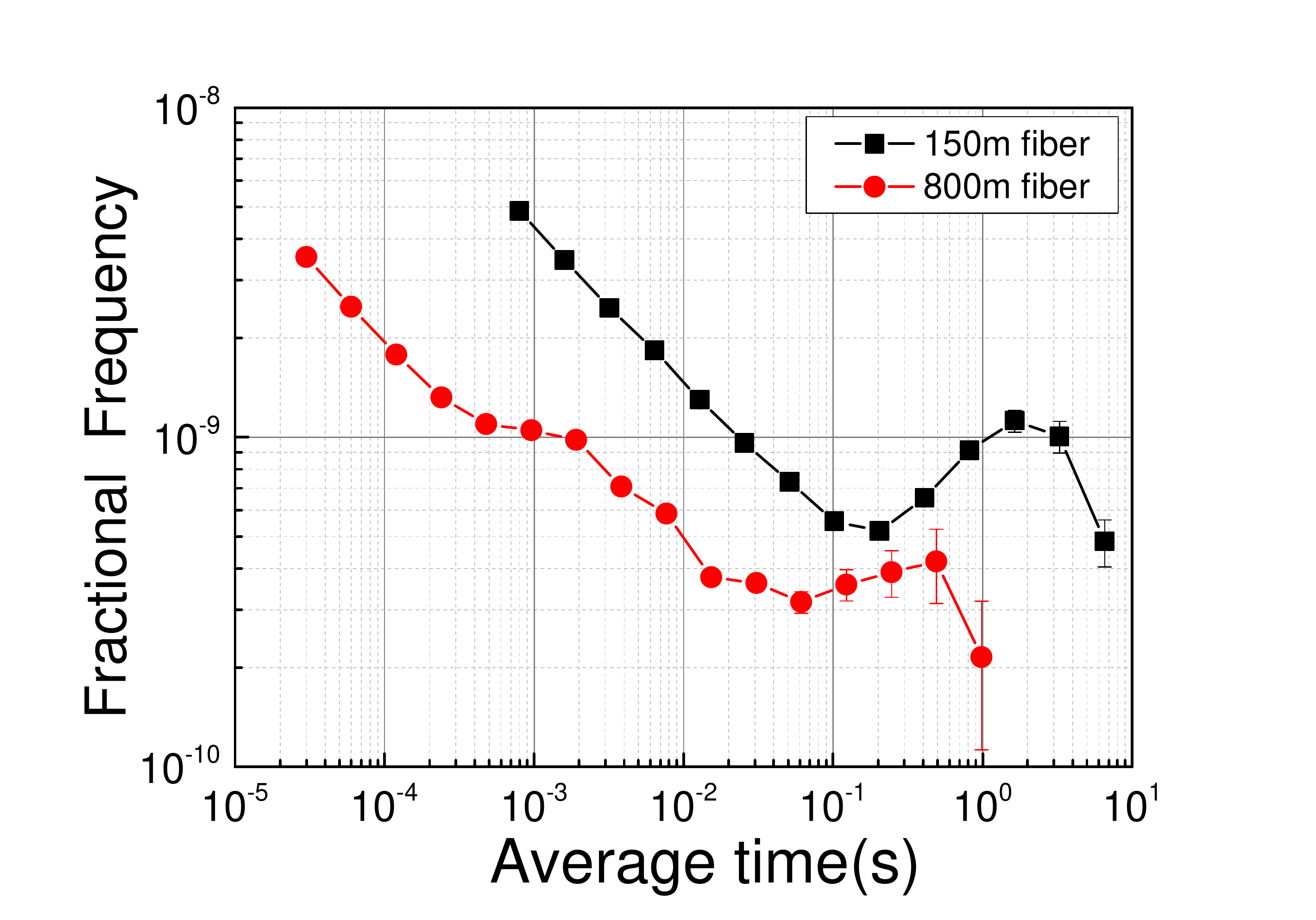}
 \caption{Fractional frequency Allan deviation of Faraday optical frequency standard}
 \label{FF}
\end{figure}

Compared with the concept of active Faraday optical clock proposed and demonstrated recently\cite{activefaraday}, the ultralong Faraday laser as an optical frequency standard demonstrated here is a good-cavity laser, instead of a bad-cavity laser. Currently, the long term stability of the ultralong Faraday laser is limited by cavity noise. If we use ARLD, Ti:sapphire, or dye as gain medium and very narrow optical clock transitions of lattice trapped atoms\cite{hinkley2013atomic}\cite{bloom2014optical}\cite{ushijima2014cryogenic}, letting the system operate in active optical clock configuration, the long-term stability can be improved.

Compared with the current cold atom optical lattice clocks based on atomic absorption laser spectroscopy\cite{hinkley2013atomic}\cite{bloom2014optical}\cite{ushijima2014cryogenic}, the ultralong Faraday laser as an optical frequency standard is compact and simple for practical implementation and application, since no local super-narrow linewidth laser is needed.

%With much narrower bandwidth Faraday atomic filter operating on conventional optical clock transitions of lattice trapped Ca or Sr atoms, the pumping beam is unnecessarily needed.
When Faraday atomic filter is operating on conventional optical clock transitions of lattice trapped Ca, Sr atoms with much narrower bandwidth, the pumping beam is unnecessarily needed.

The Faraday rotation angle is dependent on optical depth. When $\Delta\omega$ is near $\frac{\Gamma}{2}$, the relation between rotation angle and optical depth of trapped atoms can be simplified as: $\phi=\pi\cdot OD\cdot\frac{\Delta\omega}{\Gamma}$, where $\Delta\omega=\frac{g\mu_BB}{\hbar}$, and $OD$ stands for optical depth. When $\Delta\omega=\frac{\Gamma}{2}$, the rotation angle $\phi=OD\cdot\frac{\pi}{2}$.

Since the optical depth of Sr atoms trapped in hollow-core photonic crystal fiber can reach $2.5$\cite{Katori}, the 689 nm transition with 7.6 kHz linewidth can be used to realize ultra-narrow linewidth FADOF. Recently, optical depth of $60$ was achieved by using cold, pencil-shaped cloud of $^{87}$Rb atoms\cite{cavityless}. Moreover, it has been demonstrated that a single ion can cause phase shift of $1$ radian\cite{shift}. Hence, the Faraday rotation angle can reach about $\frac{\pi}{2}$ under sufficient optical depth with cold ion and neutral atoms. On the other hand, fiber lasers of length 75 km\cite{ultralong} and 270 km\cite{270km} were reported. They can provide ultra small FSR of sub-kHz level. We believe that trapped ions, as well as neutral atoms, can also be applied to ultralong Faraday laser. Together with narrower bandwidth FADOF and longer length fiber at 100 km level, it is expected that ultralong Faraday laser as optical frequency standard can provide much better performance in the future.

In conclusion, we conceive the ultralong Faraday laser concept, and implement a proof-of-principle system with ultra-narrow FADOF and ultralong fiber cavity. This ultralong Faraday laser can be used as an optical frequency standard, which is an active optical clock\cite{activeopticalclock}\cite{activefaraday} in good-cavity regime. The active lasing frequency of the ultralong Faraday laser is always staying at the center frequency of ultra-narrow FADOF's transition frequency, that is, the crossover transition $5^{2}S_{1/2},\ F = 2 \ \rightarrow\ 5^{2}P_{3/2},\ F^{\prime} = 2,\ 3$ of $^{87}$Rb.

Moreover, the ultralong Faraday laser demonstrated in this letter marks a unique kind of optical frequency standard independent of local super-narrow linewidth laser\cite{drever1983laser}\cite{yejun}\cite{jiang}.

We appreciate D. T. Cassidy for criticism on the conversion from intensity fluctuation to frequency fluctuation near resonance. Much of the work was supported by the National Natural Science Foundation of China (NSFC) (91436210), and International Science \& Technology Cooperation Program of China (2010DFR10900).

\end{document}